\documentclass[aps,prd,floatfix,nofootinbib]{revtex4} 

\usepackage{graphics}
\usepackage{graphicx}

\usepackage{latexsym}
\usepackage[cp866]{inputenc}
\usepackage[english]{babel}

\input epsf

\begin{document}

\title{\boldmath Evidence of the four-quark nature of $f_0(980)$
and $f_0(500)$}
\author{N.~N.~Achasov,$^{1}$\footnote{achasov@math.nsc.ru}
J.~V.~Bennett,$^{2}$\footnote{jvbennet@olemiss.edu}
A.~V.~Kiselev,$^{1,3}$\footnote{kiselev@math.nsc.ru}
E.~A.~Kozyrev,$^{4,3}$\footnote{E.A.Kozyrev@inp.nsk.su} and
G.~N.~Shestakov\,$^{1}$\footnote{shestako@math.nsc.ru}}
\affiliation{$^1$\,Laboratory of Theoretical Physics, S.~L.~Sobolev
Institute for Mathematics, 630090 Novosibirsk, Russia
\\ $^2$\ University of Mississippi, Oxford, Mississippi 38677, USA,
\\ $^3$\,Novosibirsk State University, Novosibirsk 630090 ,
Russia \\ $^4$\,Budker Institute of Nuclear Physics SB RAS,
Novosibirsk 630090 , Russia}


\begin{abstract}
There exists a great deal of concrete evidence in favor of the
exotic four-quark nature of light scalars. At the same time, the
further expansion of the area of the $q^2\bar q^2$ model validity
for light scalars on ever new processes seems extremely interesting
and important. We analyze the BESIII data on the decay $J/\psi\to
\gamma\pi^0\pi^0$ and show that the results of this high-statistics
experiment can be interpreted in favor of the four-quark nature of
light scalar mesons $f_0(980)$ and $f_0(500)$.
\end{abstract}

\maketitle

\section{INTRODUCTION}

To date, an impressive array of data on light scalar mesons has been
accumulated \cite{PDG2020}. However, the nontrivial properties of
light scalar mesons remain incompletely understood and continue to
cause controversy. In particular, there is much evidence in favor of
their four-quark $q^2\bar q^2$ structure, and it is a matter for
lively discussions. The number of publications devoted to light
scalar mesons is immense. Some understanding of how theoretical and
experimental explorations related to light scalar mesons have been
developing can be gained, for example, from the following works and
reviews \cite{PDG2020,GL60,Ja77,AT04,CCL06,KZ07,Wa13,Pe16,AS19,
AE20,AH20}.

The light scalar mesons has been studied, for example, in $\pi N$
and $KN$ scattering, $p\bar p$ annihilation, central hadronic
production, $J/\psi$, $\psi'$-, $B$-, $D$-, and $K$-meson decays,
$\gamma\gamma$ formation, and $\phi$ radiative decays
\cite{PDG2020}. The aim of our work is to obtain information on the
quark structure of the $f_0(980)$ and $f_0(500)$ resonances from the
modern data presented by the BESIII Collaboration on the radiative
decay $J/\psi\to\gamma \pi^0\pi^0$ \cite{Ban14,Ab15}. We note that
very interesting mechanisms of the $f_0(500)$, $f_0(980)$, and $a_0
(980)$ production in $J/\psi\to \gamma \pi\pi$ and $J/\psi\to\gamma
\eta\pi$ were considered recently in Refs. \cite{XMO20,SLTO20}. They
discussed the unusual properties of the above states. The authors of
Ref. \cite{XMO20} take into account contributions mediated by a
$K^+K^-$ triangle loop with a photon line attached. In Ref.
\cite{SLTO20} it is assumed that the $J/\psi\to\gamma\pi^+\pi^-$ and
$J/\psi\to\gamma\pi^0\eta$ decays come from the $J/\psi\to\phi
(\omega)\pi^+\pi^-,\rho^0\pi^0\eta$ reactions, where the $\rho^0$,
$\omega$, and $\phi$ get converted into a photon via vector meson
dominance. It was found that the probabilities of the $J/\psi
\to\gamma f_0(500)$ and $J/\psi\to \gamma f_0(980)$ decays
corresponding to the considered production mechanisms are quite
small. No attempts were made to describe the BESIII data \cite{Ab15}
in these works.

This paper is organized as follows. In Sec. II we consider the
BESIII data on the $S$- and $D$-wave $\pi^0\pi^0$ mass spectra in
the $J/\psi\to\gamma\pi^0\pi^0$ decay and demonstrate the
qualitative difference between the contributions of the $f_0(500)$
and $f_0(980)$ and higher-lying resonance states. We conclude that
the production mechanism of the light scalars $f_0(500)$ and
$f_0(980)$ is very different in comparison with that of the tensor
meson $f_2(1270)$ and heavy scalar states. We also compare the
$f_0(500)$ and $f_0(980)$ production with the case of the light
pseudoscalar mesons $\eta$ and $\eta'(958)$ in a model of the
Nambu-Jona-Lasinio type. In Sec. III we describe the shape of the
$S$-wave $\pi^0\pi^0$ mass spectrum in the $f_0(500)$ and $f_0(980)$
resonance region. The four-quark mechanism of the formation of these
states considered by us is in good agreement with the BESIII data.
Thus, from the $J/\psi\to\gamma\pi^0\pi^0$ decay we have once more
evidence in favor of the four-quark nature of the $f_0(980)$ and
$f_0(500)$ resonances.

\section{\boldmath BESIII results on $J/\psi\to\gamma\pi^0\pi^0$}

Clear indications of the unusual nature of $f_0(980)$ and $a_0(980)$
mesons were given by experiments on hadronic and radiative $J/\psi$
decays, the results of which are collected in Table I. It was found
that decays with the participation of light scalars $f_0(980)$ and $
a_0(980)$ are strongly suppressed in comparison with rather
intensive decays involving the classical tensor $q\bar q$ states
$a_2(1320)$, $f_2(1270)$, and $f'_2(1525)$ \cite{PDG2020,Ma87,Ei88,
KW89}. These facts are difficult to understand from the point of
view of the $q\bar q$ model, according to which the tensor
$a_2(1320)$, $f_2(1270)$, and $f'_2(1525)$ and scalar $a_0(980)$ and
$f_0(980)$ mesons are $P$-wave states in the usual $q\bar q$ system.
However, they can be easily qualitatively explained, which was done
in Ref. \cite{Ac98}, in terms of the four-quark structure of
$a_0(980)$ and $f_0(980)$ mesons \cite{Ja77,Wa13}, by the presence
of an additional $s\bar s$ pair in their wave functions.

\vspace*{-0.3cm}
\begin{table} [!ht] \centering \caption{Branching
fractions of $J/\psi$ decays taken mainly from the Particle Data
Group \cite{PDG2020} and also from Refs. \cite{Ma87,Ei88, KW89}.}
\label{Tab1}\vspace*{0.1cm}
\begin{tabular}{|c|c|c|c|}
 \hline
 $J/\psi\to a_2(1320)\rho$ & $f_2(1270)\omega$ & $f'_2(1525)\phi$ & $\gamma f_2(1270)$, \ $\qquad\gamma f'_2(1525)$ \\ \hline
 $(1.09\pm0.22)\%$ & $(4.3\pm0.6)\times10^{-3}$ & $(8\pm4)\times10^{-4}$ & $(1.64\pm0.12)\times10^{-3}$,\ \  $(5.7\pm^{0.8}_{0.5})\times10^{-4}$
 \\ \hline
 $J/\psi\to a_0(980)\rho$ & $f_0(980)\omega$ & $f_0(980)\phi$ & $\gamma f_0(980)$  \\ \hline
 $<4.4\times10^{-4}$\ \  \cite{Ma87,KW89} & $(1.4\pm0.5)\times10^{-4}$ & $(3.2\pm0.9)\times10^{-4}$ & $<1.4\times10^{-5}$\ \  \cite{Ei88} \\
 \hline \end{tabular}
 \end{table}

In the last decade, the BESIII Collaboration has made considerable
efforts to improve the knowledge of the scalar and tensor meson
sector with a series of the partial wave analyses of radiative $J/
\psi$ decays to $\eta\eta$ \cite{Ab13}, $\pi^0\pi^0$ \cite{Ban14,
Ab15}, $\eta\pi^0$ \cite{Ab16}, and $K_SK_S$ \cite{Ab18}. The BESIII
results on the $J/\psi\to\gamma\pi^0\pi^0$ decay that are important
for our following discussion of the dynamics of the $f_0(980)$ and
$f_0(500)$ production are shown in Fig. 1. The novelty of the
presented data consists in their incredible statistical precision
and purity provided by $(1.311\pm0.011)\times10^9$ $J/\psi$ decays
collected by the BESIII detector. According to the mass-independent
amplitude analysis \cite{Ban14, Ab15}, the full $\pi^0\pi^0$ mass
spectrum presented in Fig. 1(a) is saturated with the contributions
of the $D\ (2^{++})$ and $S\ (0^{++})$ waves shown in Figs. 1(b) and
1(c), respectively. We denote the invariant mass of the $\pi^0\pi^0$
system as $\sqrt{s}$. For our purpose, it is quite sufficient to use
the data presented in Refs. \cite{Ban14, Ab15} for one of the two
ambiguous solutions, especially since there is only one solution in
the $\sqrt{s}$ region up to 1 GeV.
\begin{figure} 
\begin{center}\includegraphics[width=12cm]{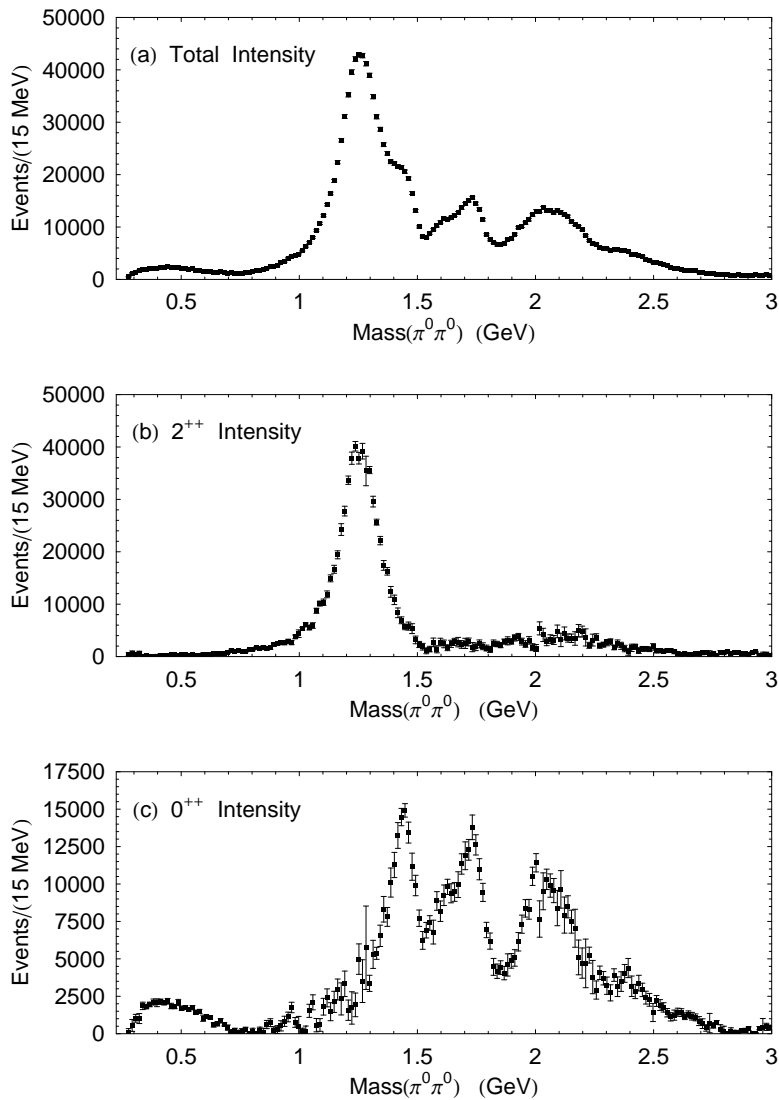}
\caption{\label{Fig1} The (a) total, (b) $D$-wave, and (c) $S$-wave
$\pi^0\pi^0$ mass spectra in the $J/\psi\to\gamma\pi^0\pi^0$ decay
presented by BESIII \cite{Ab15} for one of the two ambiguous
solutions (the first of two nominal results). There is an
unambiguous solution below 1 GeV.}\end{center}\end{figure}

The branching fraction of $J/\psi\to\gamma\pi^0\pi^0$ measured by
BESIII \cite{Ab15} is determined to be $(1.15\pm0.05)\times10^{-3}$,
where the uncertainty is systematic only and the statistical
uncertainty is negligible. Using this result at \cite{Ab15}
containing the numerical values of the $\pi^0\pi^0$ mass
distributions, we estimate the branching fractions for various
resonance type enhancements observed in $\pi^0\pi^0$ mass spectra in
the $J/\psi\to\gamma\pi^0\pi^0$ decay. The branching fractions are
\begin{eqnarray}\label{Eq1}
\mathcal{B}(J/\psi\to\gamma
f_0(500)\to\gamma\pi^0\pi^0)=\mathcal{B}(J/\psi\to\gamma\pi^0\pi^0)
N_S(2m_\pi<\sqrt{s}<0.9\ \mbox{GeV})/N_{tot}=0.324\times10^{-4},\\
\label{Eq2} \mathcal{B}(J/\psi\to\gamma
f_0(980)\to\gamma\pi^0\pi^0)=\mathcal{B}(J/\psi\to\gamma\pi^0\pi^0)
N_S(0.9\ \mbox{GeV}<\sqrt{s}<1\ \mbox{GeV})/N_{tot}=0.425\times10^{-5},\\
\label{Eq3} \mathcal{B}(J/\psi\to\gamma
f_0(1440)\to\gamma\pi^0\pi^0)=\mathcal{B}(J/\psi\to\gamma\pi^0\pi^0)
N_S(1\ \mbox{GeV}<\sqrt{s}<1.5\ \mbox{GeV})/N_{tot}=0.131\times10^{-3},\\
\label{Eq4} \mathcal{B}(J/\psi\to\gamma
f_0(1710)\to\gamma\pi^0\pi^0)=\mathcal{B}(J/\psi\to\gamma\pi^0\pi^0)
N_S(1.5\ \mbox{GeV}<\sqrt{s}<1.85\ \mbox{GeV})/N_{tot}=0.152\times10^{-3},\\
\label{Eq5} \mathcal{B}(J/\psi\to\gamma
f_0(2020)\to\gamma\pi^0\pi^0)=\mathcal{B}(J/\psi\to\gamma\pi^0\pi^0)
N_S(1.85\ \mbox{GeV}<\sqrt{s}<2.25\ \mbox{GeV})/N_{tot}=0.145\times10^{-3},\\
\label{Eq6} \mathcal{B}(J/\psi\to\gamma
f_2(1270)\to\gamma\pi^0\pi^0)=\mathcal{B}(J/\psi\to\gamma\pi^0\pi^0)
N_D(2m_\pi<\sqrt{s}<1.5\ \mbox{GeV})/N_{tot}=0.482\times 10^{-3},
\end{eqnarray} where $N_S$ and $N_D$ (for corresponding intervals of
$\sqrt{s}$), and $N_{tot}$ are the $S$ and $D$ wave, and total
numbers of $J/\psi\to\gamma\pi^0\pi^0$ events, respectively; three
$S$-wave enhancements above 1 GeV are conventionally labeled as
$f_0(1440)$, $f_0(1710)$, and $f_0(2020)$ in accordance with the
visible peak positions in Fig. 1(c).

Thus it is seen that the production of light scalar mesons is
strongly suppressed in comparison with the production of the
classical tensor meson $f_2(1270)$:
\begin{eqnarray}\label{Eq7}
\frac{\mathcal{B}(J/\psi\to\gamma f_0(500)\to\gamma\pi^0\pi^0)}{
\mathcal{B}(J/\psi\to\gamma f_2(1270)\to\gamma\pi^0\pi^0)}\approx
0.067,\qquad \frac{\mathcal{B}(J/\psi\to\gamma f_0(980)\to\gamma
\pi^0\pi^0)}{ \mathcal{B}(J/\psi\to\gamma f_2(1270)\to\gamma \pi^0
\pi^0)}\approx0.0088.\end{eqnarray} Naturally, this may indicate a
fundamental difference in the corresponding production mechanisms.

The $f_0(500)$ and $f_0(980)$ production is also significantly
suppressed in comparison with the production of heavy scalar states.
The $f_0(500)-f_0(980)$ resonance complex at $\sqrt{s}<2m_{K^+}$ is
coupled only with the $\pi\pi$ decay channel. The $f_2(1270)$
resonance is also practically elastic, $\mathcal{B}(f_2(1270)\to
\pi\pi)\approx85\%$ \cite{PDG2020}. If the $S$-wave peak in the
vicinity of 1440 MeV is related to the complex of conventional
$f_0(1370)$ and $f_0 (1500)$ resonances \cite{PDG2020}, then with
taking into account their significant inelasticity \cite{PDG2020}
and  Eqs. (3) and (6), it cannot be ruled out that the full
branching fraction $\mathcal{B}(J/\psi\to\gamma f_0(1440))
\approx\mathcal{B}(J/\psi \to\gamma f_2(1270))$, i.e., much larger
than $\mathcal{B}(J/\psi \to\gamma (f_0(500))$ and, certainly,
$\mathcal{B}(J/\psi\to\gamma (f_0(980))$. The data on the
$f_0(1710)$ resonance \cite{PDG2020,Ab18} and Eq. (\ref{Eq4}) also
support a similar proportion. It is likely that the $f_0(2020)$
resonance, which needs confirmation, is also very inelastic
\cite{PDG2020}.

Thus, one can conclude that the production mechanism of the light
scalars $f_0(500)$ and $f_0(980)$ is very different in comparison
with that of the tensor meson $f_2(1270)$ and heavy scalar states.

Note also that in a chiral-symmetric model of the Nambu-Jona-Lasinio
type \cite{VR06} it would be quite natural to expect the following
approximate relations between the decay probabilities: $\mathcal{B}
(J/\psi\to\gamma\eta)\approx\mathcal{B}(J/\psi\to\gamma f_0(500))$
and $\mathcal{B}(J/\psi\to\gamma\eta'(958))\approx\mathcal{B}(J/\psi
\to\gamma f_0(980))$. However, this is not the case. If $\mathcal{B}
(J/\psi\to\gamma f_0(500))\approx0.97\times10^{-4}$ and
$\mathcal{B}(J/\psi\to\gamma f_0(980))\approx1.27\times10^{-5}$
according to Eqs. (\ref{Eq1}) and (\ref{Eq2}), then according to the
Particle Data Group \cite{PDG2020}, $\mathcal{B}(J/
\psi\to\gamma\eta)\approx 1.11 \times10^{-3}$ and
$\mathcal{B}(J/\psi\to\gamma \eta'(958))\approx 5.25 \times10^{-3}$.
This fact is also a clear confirmation of the unusual internal
structure of the $f_0(500)$ and $f_0(980)$ resonances in comparison
with the typical $q\bar q$ mesons $\eta$ and $\eta'(958)$.
Considerations related to the chiral-symmetric model of the
Nambu-Jona-Lasinio type were used previously in Ref. \cite{AK12}.


\section{\boldmath Description of the $f_0(500)$ and $f_0(980)$ resonance
region.}

We now turn to description of the shape of the $S$-wave $\pi^0\pi^0$
mass spectrum in the $f_0(500)$ and $f_0(980)$ resonance region (see
Fig. \ref{Fig3} below). This spectrum is defined by only one
invariant amplitude $F_{J/\psi\to\gamma\pi^0\pi^0}(s)$ and can be
written in the form
\begin{eqnarray}\label{Eq7}
\frac{dN_S(s)}{d\sqrt{s}}=\frac{1}{12\pi}\,|F_{J/\psi\to\gamma\pi^0\pi^0}
(s)|^2\,k^3_\gamma(s)\,\frac{\rho_{\pi\pi}(s)}{32\pi}\,\frac{2\sqrt{s}
}{\pi},\end{eqnarray} where
$k_\gamma(s)=(m^2_{J/\psi}-s)/(2m_{J/\psi})$ and $\rho_{\pi
\pi}(s)=\sqrt{1-4m^2_\pi/s}$.

To determine $F_{J/\psi\to\gamma \pi^0\pi^0}(s)$, we use the
following consideration about the dynamics of the $f_0(500)$ and
$f_0(9 80)$ resonance production. In a series of papers
\cite{AK12,A20,AKS20}, it has been shown that semileptonic decays of
$D$ and $B$ mesons can be used as probes of constituent $q\bar q$
components in the wave functions of light scalar mesons decaying
into pseudoscalar meson pairs. In particular, in Ref. \cite{AKS20}
it was demonstrated that the model according to which the $f_0(500)$
and $f_0(980)$ production proceed via direct $q\bar q\to f_0(500),
f_0(980)$ transitions ($D^+\to d\bar d\,e^+\nu_e \to[f_0(500)+f_0
(980)]e^+\nu_e\to\pi^+\pi^-e^+\nu_e$ and $D^+_s\to s\bar s\,e^+
\nu_e\to[f_0(500)+f_0(980)]e^+\nu_e \to\pi^+\pi^-e^+\nu_e$) cannot
describe simultaneously the BESIII \cite{Ab19} and CLEO
\cite{CLEO09} data on the decays $D^+\to\pi^+\pi^-e^+\nu_e$ and
$D^+_s\to\pi^+\pi^-e^+\nu_e$. Figuratively, one can say that the
$q\bar q$ probe existing in semileptonic $(D^+,D^+_s)\to\pi^+\pi^-
e^+\nu_e$ decays does not find, to a first approximation, the
corresponding $q\bar q$ components in $f_0(500)$ and $f_0(980)$
mesons. However, the $f_0(500)-f_0(980)$ resonance complex can be
produced via seed four-quark fluctuations $q\bar q\to\pi\pi$ and
$q\bar q\to K\bar K$, which are then dressed by strong interactions
in the final state. This four-quark production mechanism provide a
good description of all details of the $\pi^+\pi^-$ mass spectra
measured in above mentioned experiments \cite{AKS20}.

In the case of the radiative $J/\psi\to\gamma\pi^0\pi^0$ decay, the
role of the $q\bar q$ probe is played by the transition
$J/\psi\to\gamma c\bar c\to\gamma gg\to\gamma q\bar q$, where $g$ is
a gluon. This source generates the classical $q\bar q$ resonance
$f_2(1270)$ and its scalar $P$-wave partners in the $q\bar q$
multiplet. However, it does not work in the case of light scalar
resonances. We assume that the $f_0(500)$ and $f_0(980)$ states are
created in the radiative $J/\psi$ decay due to all possible seed
four-quark fluctuations $gg\to q\bar qq\bar q\to\pi\pi,K\bar K$. The
described picture of the creation of light four-quark resonances in
$J/\psi\to \gamma\pi^0\pi^0$ can be effectively realized in the
language of hadronic states, as this is shown diagrammatically in
Fig. 2.
\begin{figure} [!ht] 
\begin{center}\includegraphics[width=13cm]{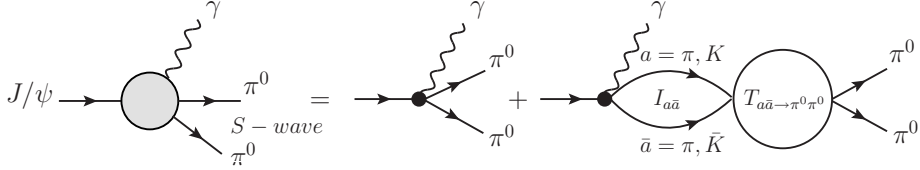}
\caption{\label{Fig2} Production of the $f_0(500)-f_0(980)$
resonance complex in $J/\psi\to\gamma\pi^0\pi^0$.}
\end{center}\end{figure}

According to this figure, we write the amplitude $F_{J/\psi\to
\gamma\pi^0\pi^0}(s)$ from Eq. (\ref{Eq7}) in the form
\begin{eqnarray}\label{Eq8}
F_{J/\psi\to\gamma\pi^0\pi^0}(s)=\lambda_{\pi\pi}\left[1+I_{\pi^+
\pi^-}(s)\,T^0_0(s)\right]+\lambda_{K\bar K}\left[
I_{K^+K^-}(s)+I_{K^0\bar K^0}(s)\right]\,T_{K^+\bar K^+\to\pi^0
\pi^0}(s),\end{eqnarray} where
$T^0_0(s)=T_{\pi^+\pi^-\to\pi^0\pi^0}(s)+\frac{1}{2}\,T_{\pi^0\pi^0
\to\pi^0\pi^0}(s)$ is the $S$-wave amplitude of the reaction
$\pi\pi\to\pi\pi$ in the channel with isospin $I=0$ composed of the
amplitudes related to individual charge channels;
$T^0_0(s)=[\eta^0_0(s)\exp(2i\delta^0_0(s))-1]/(2i\rho_{\pi\pi}(s))$,
where $\eta^0_0(s)$ and $\delta^0_0(s)$ are the corresponding
inelasticity and phase of $\pi\pi$ scattering \cite{AK06};
$T_{K^+K^-\to\pi^0 \pi^0}(s)$ is the amplitude of the $S$-wave
transition $K^+K^-\to\pi^0\pi^0$; $T_{K^+ K^-\to\pi^0\pi^0}
(s)=T_{K^0\bar K^0\to\pi^0\pi^0}(s)$ \cite{AK06}. Functions
$I_{a\bar a}(s)$ (where $a\bar a=\pi^+\pi^-,\pi^0\pi^0,
K^+K^-,K^0\bar K^0$) are the amplitudes of the loop diagrams
describing $a\bar a\to a\bar a\to$({\it the scalar state with a
virtual mass equaling} $\sqrt{s}$) transitions in which initial
$a\bar a$ pairs are produced by the underlain gluon source, $gg\to
q\bar qq\bar q\to a\bar a$, described by coupling constants
$\lambda_{a\bar a}$. Above the $a\bar a$ threshold, $I_{a\bar a}(s)$
has the form \cite{AK06}
\begin{eqnarray}\label{Eq9}
I_{a\bar a}(s)=\tilde{C}_{a\bar a}+\rho_{a\bar
a}(s)\left(i-\frac{1}{\pi}\ln\frac{1+\rho_{a\bar a}(s)
}{1-\rho_{a\bar a}(s)}\right),\end{eqnarray} where $\rho_{a\bar
a}(s)=\sqrt{1-4m^2_a/s}$ (we put $m_{\pi}\equiv m_{\pi^0}=m_{\pi^+}$
and take into account the mass difference of $K^+$ and $K^0$); if
$\sqrt{s}<2m_K$, then $\rho_{K\bar K}(s)\to i|\rho_{K\bar K}(s)|$;
$\tilde{C}_{\pi^+\pi^-}=\tilde{C}_{\pi^0\pi^0}$ and
$\tilde{C}_{K^+K^-}=\tilde{C}_{K^0 \bar K^0}$ are subtraction
constants in the loops.

We take the amplitudes $T^0_0(s)$ and $T_{K\bar K\to\pi\pi}(s)=T_{
\pi\pi\to K\bar K}(s)$ from Ref. \cite{AK06} (corresponding to
fitting variant 1 for parameters from Table 1 therein). Note that
our principal conclusions are independent of a concrete fitting
variants presented in Refs. \cite{AK06,AK11,AK12a}, containing the
excellent simultaneous descriptions of the phase shifts,
inelasticity, and mass distributions in the reactions
$\pi\pi\to\pi\pi$, $\pi\pi\to K\bar K$, and $\phi\to\pi^0\pi^0
\gamma$. The amplitudes $T^0_0(s)$ and $T_{\pi\pi\to K\bar K}(s)$
were described in Refs. \cite{AK06,AK11,AK12a} by the complex of the
mixed $f_0(500)$ and $f_0(980)$ resonances and smooth background
contributions. The constructed $\pi\pi$ scattering amplitude
$T^0_0$(s) \cite{AK06, AK11,AK12a} has regular analytical properties
in the $s$ complex plane and describes both experimental data and
the results based on chiral expansion and Roy equations
\cite{CCL06}.

Note that the phase of the amplitude $F_{J/\psi\to\gamma\pi^0\pi^0}
(s)$ in Eq. (\ref{Eq8}) [taking into account Eq. (\ref{Eq9})]
coincide with the $\pi\pi$ scattering phase $\delta^0_0(s)$ below
the $K^+K^-$ threshold where $\eta^0_0(s)=1$ [as is the phase of the
amplitude $T_{K^+ K^-\to\pi^0\pi^0}(s)$ \cite{AK06}].

For reasons of $SU(3)$ symmetry, we will assume that the seed
coupling constants in Eq. (\ref{Eq8}) are the same: $\lambda_{\pi
\pi}=\lambda_{K\bar K}\equiv\lambda$. Thus, $\lambda$ defines the
overall normalization. Since the amplitudes $T^0_0(s)$ and
$T_{K^+K^-\to\pi^0\pi^0}(s)$ are known \cite{AK06} from the analysis
of the data on the reactions $\pi\pi\to\pi\pi$, $\pi\pi\to K\bar K$,
and $\phi\to\pi^0\pi^0 \gamma$, then we have only two parameters
$\tilde{C}_{\pi^+\pi^-}$ and $\tilde{C}_{K^+K^-}$ to describe the
shape of the $\pi^0\pi^0$ mass spectrum.

\begin{figure} [!ht] 
\begin{center}\includegraphics[width=12cm]{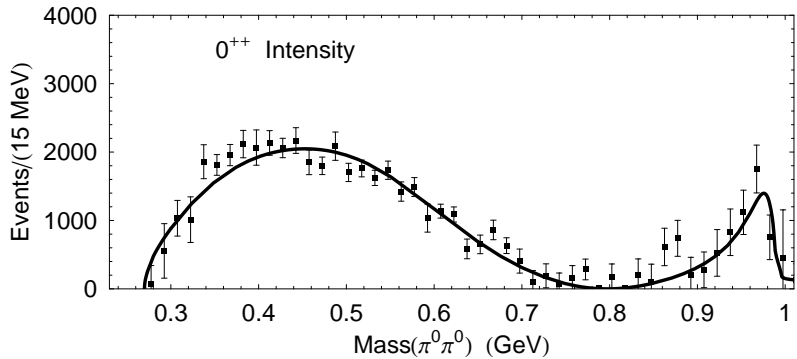}
\caption{\label{Fig3} The $S$-wave $\pi^0\pi^0$ mass spectrum in the
$f_0(500)$ and $f_0(980)$ resonance region in the $J/\psi\to\gamma
\pi^0\pi^0$ decay. The points with the error bars are the BESIII
data \cite{Ab15}. The solid curve represents our fit.}\end{center}
\end{figure}

The choice of $\tilde{C}_{\pi^+\pi^-}=1.052$ and $\tilde{C}_{K^+K^-}
=0.2396$ provides a good description of the BESIII data \cite{Ab15}
on the $S$-wave $\pi^0\pi^0$ mass spectrum in the $J/\psi\to\gamma
\pi^0\pi^0$ decay at $\sqrt{s}<1$ GeV, see Fig. \ref{Fig3}. The
curve in Fig. 3 corresponds to $\chi^2/(n.d.f)\approx57.4/(49-3)
\approx1.25$. In so doing the relative errors of the fitting
parameters $\lambda$, $\tilde{C}_{\pi^+\pi^-}$, and
$\tilde{C}_{K^+K^-}$ are approximately $\pm1.9\%$, $\pm2.4\%$, and
$\pm9.3\%$, respectively. Thus, we obtained once more evidence in
favor of the four-quark nature of the $f_0(980)$ and $f_0(500)$
resonances. Their production occurs due to four-quark transitions.
It is interesting to compare the resulting picture on the
$\pi^0\pi^0$ mass spectrum with the $S$-wave $\pi\pi$ elastic cross
section. This cross section reaches the unitary limit in the energy
range where the $S$-wave $\pi^0\pi^0 $ mass spectrum in
$J/\psi\to\gamma \pi^0\pi^0$ has practically zero minimum. It has
also a narrow dip at the site of the $f_0(980)$ resonance (compare
with Fig. \ref{Fig3}). The $S$-wave $\pi\pi$ elastic cross section
falls naturally in the inelastic region ($\sqrt{s}>1$ GeV) with
increasing energy, in contrast to the $S$-wave $\pi^0\pi^0$ mass
spectrum in the $J/\psi\to\gamma \pi^0\pi^0$ decay [see Fig. 1(c)].

As we have already noted in the Introduction, the nontrivial
properties of light scalar mesons remain not fully understood and
continue to cause controversy. Different modern points of view on
the quark structure of the $f_0(980)$ and $f_0(500)$ mesons are
remarkably presented in the minireviews ``Scalar mesons below 2
GeV'' \cite{AE20} and ``Non-$q\bar q$ states'' \cite{AH20}. We hope
that our evidence will serve the general desire to better understand
the physics of light scalars.\\ 

\begin{center} {\bf ACKNOWLEDGMENTS} \end{center}

The work of N. N. A., A. V. K., and G. N. S. was carried out within
the framework of the state contract of the Sobolev Institute of
Mathematics, Project No. 0314-2019-0021.


\begin{thebibliography}{99}
\bibitem{PDG2020} P. A. Zyla {\it et al.} (Particle Data Group), Prog. Theor. Exp. Phys. {\bf 2020}, 083C01 (2020).
\bibitem{GL60}  M. Gell-Mann and M. Levy, Nuovo Cimento {\bf 16}, 705 (1960).
\bibitem{Ja77}  R. L. Jaffe, Phys. Rev. D {\bf 15}, 267 (1977); {\bf 15}, 281 (1977).
\bibitem{AT04}  C. Amsler and N. A. Tornqvist, Phys. Rep. 389, 61 (2004).
\bibitem{CCL06} I. Caprini, G. Colangelo, and H. Leutwyler, Phys. Rev. Lett. {\bf 96}, 132001 (2006).
\bibitem{KZ07}  E. Klempt and A. Zaitsev, Phys. Rep. 454, 1 (2007).
\bibitem{Wa13}  S. Weinberg, Phys. Rev. Lett. {\bf 110}, 261601 (2013).
\bibitem{Pe16}  J. R. Pelaez, Phys. Rep. {\bf 658}, 1 (2016).
\bibitem{AS19}  N. N. Achasov and G. N. Shestakov, Usp. Fiz. Nauk, {\bf 189}, 3 (2019) [Phys. Usp. {\bf 62}, 3 (2019)].
\bibitem{AE20}  C. Amsler, S. Eidelman, T. Gutsche, C. Hanhart, and S. Spanier, Scalar mesons below 2 GeV, Mini-review in Ref. \cite{PDG2020}.
\bibitem{AH20}  C. Amsler and C Hanhart, Non-$q\bar q$ states, Mini-review in Ref. \cite{PDG2020}.
\bibitem{Ban14} J. V. Bennett, Ph. D. thesis, Indiana University, 2014.                                 
\bibitem{Ab15}  M. Ablikim {\it et al.} (BESIII Collaboration), Phys. Rev. D {\bf 92}, 052003 (2015).   
\bibitem{XMO20} C. W. Xiao, U.-G. Meissner, and J. A. Oller, Eur. Phys. J. A {\bf 56}, 23 (2020).
\bibitem{SLTO20}S. Sakai, W. H. Liang, G. Toledo, and E. Oset, Phys. Rev. D {\bf 101}, 014005 (2020).
\bibitem{Ma87}  U. Mallik, SLAC Report No. SLAC-PUB-4238, Stanford, 1987.
\bibitem{Ei88}  G. Eigen, in {\it Proceedings of the XXIV International Conference on High Energy Physics},
                edited by R. Kotthaus and J. H. K\"{u}hn (Springer-Verlag, Munich, 1988), Session 4, p. 590.
\bibitem{KW89}  L. K\"{o}pke and N. Wermes, Phys. Rep. {\bf 174}, 67 (1989).
\bibitem{Ac98}  N. N. Achasov, Usp. Fiz. Nauk, {\bf 168}, 1257 (1998) [Phys. Usp. {\bf 41}, 1149 (1998)].
\bibitem{Ab13}  M. Ablikim {\it et al.} (BESIII Collaboration), Phys. Rev. D {\bf 87}, 092009 (2013).   
\bibitem{Ab16}  M. Ablikim {\it et al.} (BESIII Collaboration), Phys. Rev. D {\bf 94}, 072005 (2016).   
\bibitem{Ab18}  M. Ablikim {\it et al.} (BESIII Collaboration), Phys. Rev. D {\bf 92}, 072003 (2018).   
\bibitem{VR06}  M. K. Volkov and A. E. Radzhabov, Usp. Fiz. Nauk, {\bf 176}, 569 (2006) [Phys. Usp. {\bf 49}, 551 (2006)].
\bibitem{AK12} N. N. Achasov and A. V. Kiselev, Phys. Rev. D {\bf 86}, 114010 (2012).
\bibitem{A20}  N. N. Achasov, Phys. Part. Nucl., {\bf 51}, 632 (2020). 
\bibitem{AKS20}N. N. Achasov, A. V. Kiselev, and G. N. Shestakov, Phys. Rev. D {\bf 102}, 016022 (2020).
\bibitem{Ab19}   M. Ablikim {\it et al.} (BESIII Collaboration), Phys. Rev. Lett. {\bf 122}, 062001 (2019).
\bibitem{CLEO09} K. M. Ecklund {\it et al.} (CLEO Collaboration), Phys. Rev. D {\bf 80}, 052009 (2009).
\bibitem{AK06}  N. N. Achasov and A. V. Kiselev, Phys. Rev. D {\bf 73}, 054029 (2006).
\bibitem{AK11}  N. N. Achasov and A. V. Kiselev, Phys. Rev. D {\bf 83}, 054008 (2011).
\bibitem{AK12a} N. N. Achasov and A. V. Kiselev, Phys. Rev. D {\bf 85}, 094016 (2012).
\end{thebibliography}
\end{document}